\documentclass[useAMS,usenatbib]{mnras}
\usepackage{graphicx}
\usepackage{longtable}
\usepackage{amssymb}
\usepackage{amsmath}
\title[Heavy obscuration in Hot Dust-Obscured Galaxies]{Heavy X-ray obscuration in the most-luminous galaxies discovered by \textit{WISE} }
\author[F. Vito et al.]
{F. Vito,$^{1,2}$\thanks{E-mail: fvito@psu.edu}
W.N. Brandt,$^{1,2,3}$
D. Stern,$^{4}$
R.J. Assef,$^{5}$
C.-T. J. Chen,$^{1,2}$
M. Brightman,$^{6}$
\newauthor
A. Comastri,$^{7}$
P. Eisenhardt,$^{4}$
G.P. Garmire,$^{8}$
R. Hickox,$^{9}$
G. Lansbury,$^{10}$
\newauthor
C.-W. Tsai,$^{11}$
D.J. Walton,$^{10}$
J. W. Wu$^{12}$\\
$^{1}$ Department of Astronomy \& Astrophysics, 525 Davey Lab, The Pennsylvania State University, University Park, PA 16802, USA\\
$^{2}$ Institute for Gravitation and the Cosmos, The Pennsylvania State University, University Park, PA 16802, USA\\
$^{3}$ Department of Physics, The Pennsylvania State University, University Park, PA 16802, USA\\
$^{4}$ Jet Propulsion Laboratory, California Institute of Technology, 4800 Oak Grove Dr., Pasadena, CA 91109, USA\\
$^{5}$ N\'ucleo de Astronom\'ia de la Facultad de Ingenier\'ia y Ciencias, Universidad Diego Portales, Av. Ej\'ercito Libertador 441, Santiago, Chile\\
$^{6}$ Cahill Center for Astrophysics, 1216 East California Boulevard, California Institute of Technology, Pasadena, CA 91125, USA\\
$^{7}$ INAF – Osservatorio Astronomico di Bologna, Via Gobetti 93/3, I-40129 Bologna, Italy\\
$^{8}$ Huntingdon Institute for X-ray Astronomy, LLC, 10677 Franks Road, Huntingdon, PA 16652, USA\\
$^{9}$ Department of Physics and Astronomy, Dartmouth College, Hanover, NH 03755, USA\\
$^{10}$ Institute of Astronomy, University of Cambridge, Madingley Road, Cambridge, CB3 0HA, UK\\
$^{11}$ Department of Physics and Astronomy, University of California, 430 Portola Plaza, Los Angeles, CA 90095-1547, USA\\
$^{12}$ National Astronomical Observatories, Chinese Academy of Sciences, 20A Datun Road, Chaoyang District, Beijing, 100012, China\\
}

\newcommand*{\lbol}{\ensuremath{L_{\mathrm{bol}}}}

\newcommand*{\lunit}{\ensuremath{\mathrm{erg\,s^{-1}}}}
\newcommand*{\funit}{\ensuremath{\mathrm{erg\, cm^{-2}\,s^{-1}}}}
\newcommand*{\xmm}{\textit{\mbox{XMM-Newton}}}
\newcommand*{\chandra}{\textit{Chandra}}

\newcommand*{\nhunits}{\mathrm{cm^{-2}}}

\defcitealias{Hsu14}{H14}

\date{Accepted 2017 November 29 . Received 2017 November 29 ; in original form 2017 October 11}

\pubyear{2018}

\begin{document}

\graphicspath{{.}}
\pagerange{\pageref{firstpage}--\pageref{lastpage}} \pubyear{2018}
\maketitle
\label{firstpage}

\begin{abstract}
Hot Dust-Obscured Galaxies (Hot DOGs) are hyperluminous  (\mbox{$L_{\mathrm{8-1000\,\mu m}}>10^{13}\,\mathrm{L_\odot}$}) infrared galaxies with extremely high (up to hundreds of K) dust temperatures. The sources powering both their extremely high luminosities and dust temperatures are thought to be deeply buried and rapidly accreting supermassive black holes (SMBHs). Hot DOGs could therefore represent a key evolutionary phase in which the SMBH growth peaks. X-ray observations can be used to study their obscuration levels and luminosities. In this work, we present the X-ray properties of the 20 most-luminous ($L_{\mathrm{bol}}\gtrsim10^{14}\, L_\odot$) known Hot DOGs  at $z=2-4.6$. Five of them are covered by long-exposure (10--70 ks) \chandra\, and \xmm\, observations, with three being X-ray detected, and we study their individual properties.
One of these sources (W0116--0505) is a Compton-thick candidate, with column density $N_H=(1.0-1.5)\times10^{24}\,\nhunits$ derived from X-ray spectral fitting. The remaining 15 Hot DOGs have been targeted by a \textit{Chandra} snapshot (3.1 ks) survey. None of these 15 is individually detected; therefore we applied a stacking analysis to investigate their average emission. From hardness-ratio analysis, we constrained the average obscuring column density and intrinsic luminosity to be log$N_H\,\mathrm{[\nhunits]}>23.5$ and $L_X\gtrsim10^{44}\,\lunit$, which are consistent with results for individually detected sources. We also investigated the $L_X-L_{6\mu\mathrm{m}}$ and $L_X-L_{bol}$ relations, finding hints that Hot DOGs are typically X-ray weaker than expected, although larger samples of luminous obscured QSOs are needed to derive solid conclusions.

\end{abstract}

\section{Introduction}\label{intro}
According to a widely discussed framework of supermassive black hole (SMBH) and galaxy co-evolution \mbox{\citep[e.g.,][]{Hopkins08, Alexander12}}, the most-luminous \mbox{($\lbol\gtrsim10^{47}\lunit$)} quasars (QSOs), typically with black hole masses of \mbox{$\gtrsim10^9\,\mathrm{M_\odot}$}, are expected to be triggered by major ``wet" (i.e. gas-rich) mergers. The large gas and dust reservoirs feed both high-rate accretion, close to or even exceeding the Eddington limit, onto SMBHs, and starburst events in the host galaxies. This phase of extremely fast SMBH growth is hidden by the same accreting and star-forming material, which absorbs the optical-to-X-ray radiation emitted in the nuclear region, and reprocesses it into the infrared (IR) bands. Then, during the so-called ``blow-out phase", radiative feedback mechanisms sweep away the material surrounding the SMBHs, which eventually shine as blue, unobscured QSOs.

Recently, a population of rare hyper-luminous (\mbox{$L_{\mathrm{8-1000\,\mu \mathrm{m}}}>10^{13}\,\mathrm{L_\odot}$}) infrared galaxies (HyLIRGs) at $z=1.6-4.6$ has been discovered in the \textit{Wide-Field Infrared Survey Explorer} (\textit{WISE}, \citealt{Wright10}) all-sky survey \citep{Eisenhardt12, Wu12,Tsai15, Assef16}. Unlike other HyLIRGs, these objects have been selected using the ``W1W2-dropout" technique, i.e. requiring no or marginal detection in the W1 and W2 \textit{WISE} bands (3.4 and 4.6~$\mu \mathrm{m}$, respectively), and strong detections in the W3 and W4 bands (12 and 22~$\mu \mathrm{m}$, respectively). The selected objects are similar to Dust-Obscured Galaxies (DOGs; \citealt{Dey08}), but with higher dust temperatures (up to hundreds vs. $30-40$ K; e.g. \citealt{Pope08, Melbourne12, Wu12,Jones14,Tsai15}), and are therefore named Hot DOGs. 

The most probable sources powering such extreme luminosities and high dust temperature are deeply buried, high-mass, and rapidly accreting SMBHs. Evidence for the presence of obscured QSOs in these objects comes from optical spectroscopy, which often reveals QSO-like emission lines \citep{Eisenhardt12, Wu12,Tsai15, Stern14,Assef16}. Moreover, decomposition analyses of their spectral energy distributions (SEDs) require QSO emission to explain the extreme luminosities and high dust temperatures \citep{Assef15, Tsai15}. Hot DOGs therefore represent excellent candidates for high-mass heavily obscured QSOs caught during the peak of their post-merger accretion phases, when the bolometric luminosity is dominated by dust-reprocessed thermal emission. Intriguingly, these objects challenge the ``receding torus" model \citep[first suggested by][]{Lawrence91}, according to which the covering factor of the torus-shaped screen obscuring Active Galactic Nuclei (AGN) emission decreases with increasing AGN luminosity \citep[but see][]{Mateos17}. This suggests a more complex distribution of the obscuring material, perhaps distributed on scales larger than the classical torus \citep[e.g.,][]{Blecha17}. The same conclusion is reached considering Hot DOG optical spectra, which span from typical $\mathrm{Ly}\alpha$ emitters (LAEs) to broad-line AGN Types \citep{Eisenhardt12, Wu12,Tsai15, Stern14,Assef16}. Hot DOGs have space densities close to those of Type 1 QSOs with similar luminosity \citep{Assef15}; i.e., the objects populating the very tip of the optical luminosity function of QSOs, thus representing the most luminous (unlensed) optical QSOs known.

X-ray emission can be used to identify directly accreting SMBHs even when these are buried in large column densities of obscuring material, thanks to its high penetrating power \citep[e.g.,][]{Brandt15}, and it can therefore be used to investigate the basic physical properties, in terms of obscuration and luminosity, of the central engines powering Hot DOGs. A few of these extreme objects have been covered by sensitive X-ray observations. \cite{Stern14} reported that \textit{XMM-Newton} and \mbox{\textit{NuSTAR}} observations of three Hot DOGs at $z=2.1-2.4$ are consistent with these sources being heavily obscured ($N_H\gtrsim10^{24}\nhunits$). \cite{Piconcelli15} analyzed an \mbox{\textit{XMM-Newton}} spectrum of a Hot DOG at $z=2.298$, and concluded it is reflection-dominated due to Compton-thick (i.e. $N_H>1.5\times10^{24}\,\mathrm{cm^{-2}}$) absorption. \cite{Assef16} derived a heavy obscuration level ($N_H=6.3^{+8.1}_{-2.1}\times10^{23}\,\nhunits$) for a Hot DOG serendipitously observed with \chandra. \cite{RicciC17} applied a modified selection technique to \textit{WISE} all-sky survey data capable of identifying Hot-DOG  analogs at $z\approx1$. Spectral analysis of one of these objects targeted by \textit{NuSTAR} and \textit{Swift} revealed heavy obscuration ($N_H=7.1^{+8.1}_{-5.1}\times10^{23}\,\nhunits$).  However, the limited number of Hot DOGs observed in \mbox{X-rays} is a major limitation to our knowledge of the typical X-ray properties of this class of object.

In this work, we present the complete X-ray coverage of the 20 most-luminous known Hot DOGs at $z=2-4.6$. Nineteen of them are included in the \cite{Tsai15} compilation of extremely luminous ($L_{\mathrm{bol}}>10^{14}\, L_\odot$) infrared galaxies. The remaining target was not included in the \cite{Tsai15} sample, as its luminosity is slightly lower ($L_{\mathrm{bol}}=9.8\times10^{13}\, L_\odot$) than the threshold used in that work.
Our sources are among the most-luminous galaxies in the Universe, comparable with the most-luminous optically selected blue QSOs \citep[e.g.,][]{Just07}. They have been selected on the basis of their WISE colours, unlike other samples of red QSOs  with similar luminosities \citep[e.g.,][]{Urrutia05, Glikman07, Banerji15,Hamann17}, whose selections require bright optical/NIR counterparts and/or broad emission lines in optical/near-IR spectra. Hence, our sources are redder over the rest-frame $1-5 \mu\mathrm{m}$ band (see, e.g., Fig.~16 of \citealt{Hamann17}) and more representative of the early heavily obscured accretion phase corresponding to the peak of SMBH growth, during which radiative feedback from the accreting SMBHs starts sweeping away the obscuring material \citep[e.g.][]{Blecha17}. 

Four Hot DOGs in our sample were targeted by long-exposure \chandra\, and \xmm\, pointings, one is serendipitously detected in a \chandra\, observation, and the remaining fifteen were the targets of a \chandra\, snapshot (3 ks) survey.
The goal of our \chandra\, snapshot strategy is to gather a statistically significant sample of Hot DOGs covered by economical, but sensitive, X-ray observations. This allows us to make basic statements about individual targets, thus estimating the feasibility of possible future deeper \mbox{X-ray} follow-up observations aimed at performing accurate spectral analysis. Average properties (e.g. column density and luminosity) are derived by performing a stacking analysis and used together with results from individually studied objects to draw conclusions about the general properties of the Hot DOG population. Throughout this paper we use an $H_0=70\,\rmn{km\,s^{-1}Mpc^{-1}}$, $\Omega_m=0.3$, and $\Omega_\Lambda=0.7$ $\Lambda$CDM cosmology.

\section{Targeting details, X-ray observations, and data reduction}\label{obs}
In this section, we briefly describe the X-ray observations used in this work  (see Tab.~\ref{obs_tab} for details) and the data reduction. We keep separate the targets of the \chandra\, snapshot survey from those of longer X-ray observations since a stacking analysis will be performed on the former objects to derive their average properties. These combined observations provide full X-ray coverage of the \cite{Tsai15} sample of all the $L_{\mathrm{bol}}>10^{14}\, L_\odot$ known Hot DOGs.

\subsection{Targets of the \chandra\, snapshot survey}\label{snap}
We obtained \chandra\, ACIS-S snapshots (\mbox{$\approx3$ ks} each, Cycle 18) of 16 extremely luminous \mbox{($L_{\mathrm{bol}}\gtrsim10^{14}\, L_\odot$)} infrared galaxies. Fifteen of them are selected from the compilation of 20 Hot DOGs with $L_{\mathrm{bol}}>10^{14}\, L_\odot$ by \cite{Tsai15}. However, in this work we do not include one of these 15 \mbox{(W0126--0529)}\footnote{Nonetheless, we analyzed the X-ray data for this source, finding that it is not detected, with only one photon detected within 2~arcsec from the \textit{WISE} position in the $0.5-7$ keV band.} as additional observations suggest it may have a much lower redshift ($z=0.832$) than the value reported by \citet[$z=2.937$]{Tsai15} and the redshifts of the other sources.
 The remaining target (W0134--2607) was not included in the \cite{Tsai15} compilation as its luminosity \mbox{($L_{\mathrm{bol}}=9.8\times10^{13}\, L_\odot$)} is slightly lower than the threshold used in that work. Each observation was reprocessed with the \textit{chandra\_repro} script using CIAO 4.9\footnote{http://cxc.harvard.edu/ciao4.9/} and CALDB 4.7.3. Since all the observations were taken in Very Faint Mode, we applied the optimal background screening by setting the option \textit{check\_vf\_pha=yes}. We also produced exposure maps of our observations with the \textit{mkexpmap} tool in the $0.5-2$ keV, $2-7$ keV, and $0.5-7$ keV bands.

We will apply in \S~\ref{stack_sec} a stacking analysis to Hot DOGs targeted by our snapshot survey.
In order to check the astrometry of the observations, whose potential problems can affect the stacking results, we created images in the $0.5-2$ keV  band,  PSF maps, and exposure-time maps (both at a single energy of 1.49~keV) using standard CIAO tools (\textit{dmcopy}, \textit{mkpsfmap}, \textit{mkinstmap} and \textit{mkexpmap}) and ran \textit{wavdetect} with a significance threshold of $10^{-6}$. For each observation, we detected up to four \mbox{X-ray} sources (with typically only 2--4 counts) on the target chip, and matched their positions with SDSS and USNO-B1 optical source catalogs. Considering the small number of detected sources and that the typical offsets are within the X-ray positional uncertainties, we did not apply any astrometric corrections. We ran the detection procedure in the $0.5-7$ keV band and obtained similar results. Relaxing the significance threshold (to $10^{-5}$) causes the detection of several sources with counts as low as one (in low-background areas), which are therefore most probably spurious. 

\subsection{Hot DOGs covered by long-exposure \mbox{X-ray} observations}\label{sample_indiv}
The five Hot DOGs from the \cite{Tsai15} compilation not included in our snapshot survey had been previously covered by \chandra\, and \xmm\, observations. Two of them were targeted with \chandra\, (W0116--0505 and W0220+0137, PI: R. Assef), two sources were targeted with \xmm\, (W1838+3429 and W2246--0526,\footnote{W2246--0526 was targeted with \xmm\, in 2015, but that observation was strongly affected by background flaring. Thus, it was observed again in 2016. We use the latter dataset.} PI: D. Stern),\footnote{The same \xmm\, program targeted a third Hot DOG, W2026+0716, which has a lower bolometric luminosity than the other galaxies we are considering in this work. Nevertheless, in Appendix \ref{appendix} we discuss its X-ray properties in the context of the results of this paper.} while the last one (W0134--2922) is serendipitously covered by both \chandra\,($\approx49$ ks) and \xmm\, ($\approx40$ and 19 ks) observations. For W0134--2922, we only use the more sensitive \chandra\, dataset. \chandra\, observations have been reduced as in \S\ref{snap}. 
W0220+0137 was observed in three separate pointings. We used the \textit{merge\_obs} tool to merge the individual observations and produce exposure maps in the same three energy bands. 

 For \xmm\, observations, we used the Science Analysis System (SAS) 16.1.0 and HEASOFT 6.21
to analyze the Observation Data Files (ODFs). 
The ODFs were processed with the SAS tasks {\sc epicproc} ({\sc epproc} and {\sc emproc} for 
PN and MOS, respectively) to create MOS1, MOS2, and PN out-of-time (OOT) event files for each ObsID. 
We removed the time intervals that were severely affected by flaring backgrounds following the standard approach suggested by the  {\it XMM-Newton} Science Operations Centre.\footnote{\url{https://www.cosmos.esa.int/web/xmm-newton/sas-thread-epic-filterbackground}}
The background-filtered exposure times for PN are 10.5~ks for W1838+3429, and 25~ks for W2246-0526.
After screening for background flares, we further excluded events in energy ranges that overlap with strong instrumental background lines (Al K$\alpha$ lines at 1.45--1.54~keV for MOS and PN; Cu lines at 7.2--7.6~keV and 7.8--8.2~keV for PN).
From the background-screened, instrumental-line-removed event files, we extract images with a $4^{\prime\prime}$ pixel size using {\sc evselect} in the  0.5--2 keV, 2--10 keV, and 0.5--10~keV bands.
The vignetting-corrected exposure maps are then generated using the SAS task {\sc eexpmap}.

\begin{table*}
	\caption{ Properties of our sample (from \citealt{Tsai15} and Eisenhardt et al., in prep.) and basic information on the X-ray observations. In the ``spectral Type" column, LAE and NLAGN stands for $Ly\alpha$ emitter and narrow-line AGN, respectively.}\label{obs_tab}
	\begin{tabular}{ccccccccc}
		\hline
		\multicolumn{1}{c}{\textit{WISE}} &
		\multicolumn{1}{c}{RA [J2000]} &
		\multicolumn{1}{c}{DEC [J2000]}& 
		\multicolumn{1}{c}{$z$}&
		\multicolumn{1}{c}{$L_{\mathrm{bol}}$ [$10^{13}\,L_\odot$]}& 
		\multicolumn{1}{c}{Spec. Type}& 
		\multicolumn{1}{c}{ObsID}& 		
		\multicolumn{1}{c}{Date}&
		\multicolumn{1}{c}{$\mathrm{T_{exp}}$ [ks]}\\
		\hline
		\multicolumn{9}{c}{Targets of the \chandra\, snapshot survey}\\
		W0134--2607 & 01:34:00.59 & --26:07:26.6 & 2.142 & 9.8 & NLAGN &  \chandra\,19743 & 2017-09-24 & 3.1\\
		W0149+2350 & 01:49:46.18 & +23:50:14.6 & 3.228 & 10.4 & AGN &  \chandra\,19729 & 2017-09-25 & 3.1\\
		W0255+3345 & 02:55:34.9 & +33:45:57.8 & 2.668 & 10.4& LAE &  \chandra\,19730& 2017-07-07 & 3.1\\
		W0410--0913  & 04:10:10.6  &  --09:13:05.2 & 3.592  &  16.8& LAE & \chandra\,19731  & 2017-02-27& 3.1\\
		W0533--3401  & 05:33:58.4 & --34:01:34.5 & 2.904& 10.4 & LAE &   \chandra\,19732 &2017-03-23 & 3.1\\
		W0615--5716  & 06:15:11.1  &  --57:16:14.6  & 3.399 &  16.5 & LAE &  \chandra\,19733  & 2017-04-03& 3.1\\
		W0831+0140 & 08:31:53.3 &  +01:40:10.8 & 3.888& 18.0 & AGN &   \chandra\,19734& 2017-04-19 & 3.1\\
		W0859+4823 & 08:59:29.9 & +48:23:02.0 &3.245 & 10.0 &AGN &  \chandra\,19735 & 2017-01-01 & 3.1\\
		W1248--2154 & 12:48:15.2  & --21:54:20.4 &  3.318 &  11.8& LAE &  \chandra\,19736 & 2017-02-05& 3.1\\
		W1322--0328 & 13:22:32.6 & --03:28:42.2 & 3.043 & 10.1&  LAE & \chandra\,19737 & 2017-03-09& 3.1\\
		W2042--3245 & 20:42:49.3 & --32:45:17.9 & 3.963 & 13.7& LAE & \chandra\,19738  & 2017-06-08& 3.1\\
		W2201+0226 & 22:01:23.39 & +02:26:21.8 & 2.877 & 11.9 & LAE & \chandra\, 19739     & 2017-09-28  & 3.1\\
		W2210--3507 & 22:10:11.9 & --35:07:20.0 &  2.814 & 11.7 & LAE &  \chandra\,19740 &  2017-04-07&  3.1\\
		W2246--7143 & 22:46:12.1 & --71:44:01.3 & 3.458  & 12.1 & LAE & \chandra\, 19741 & 2016-12-28 &3.1\\
		W2305--0039 & 23:05:25.9 & --00:39:25.7 & 3.106  & 13.9 & LAE/NLAGN & \chandra\, 19742 & 2017-01-10 & 3.1\\
		\hline
		\multicolumn{9}{c}{Targets of long-exposure X-ray observations}\\		
			W0116--0505 &01:16:01.42 & --05:05:04.2 & 3.173 &11.7 & AGN & \chandra\,18210 & 2016-09-12 & 70.1 \\
		W0134--2922$^a$ &01:34:35.71 & --29:22:45.4 & 3.047 &11.3 & AGN/NLAGN &\chandra\,16351 & 2014-08-21& 48.8  \\	
		W0220+0137$^b$ & 02:20:52.13 & +01:37:11.4  & 3.122	 & 12.9 & AGN & \chandra\,18707  & 2015-11-28& 19.8\\
		&                     &                    &            &      &     &\chandra\,18708 & 2015-11-21 &   31.3\\
		&                     &                    &            &      &     &\chandra\,18211  & 2015-11-26&   19.8\\
		W1838+3429& 18:38:09.16 & +34:29:25.9 & 3.205 & 11.1 &  LAE &\textit{XMM} 0764400301	 & 2015-09-24&10.5$^c$\\	
		W2246--0526& 22:46:07.57 & --05:26:35.0&4.593 & 34.9& AGN &\textit{XMM} 0783670101   & 2016-05-17& 25.0$^c$\\

		\hline
	\end{tabular}\\
$^a$ This source is serendipitously covered by \chandra\, and \xmm\, observations. Only the \chandra\, dataset, where the source is located at an off-axis angle of 3.6 arcmin, is used here. $^b$ For this source we used the combined (with the CIAO tool \textit{merge\_obs}) \chandra\, observations. $^c$ Filtered for background flaring. Only the $pn$ instrument is used here for simplicity.
\end{table*}

\section{Detection procedure, photometry, and results for individual sources}\label{detection}
We assessed the detection significance of our sources by computing the binomial no-source probability \citep{Weisskopf07,Broos07}
\begin{equation}\label{PB}
P_B(X\geq S)= \sum_{X=S}^{N}\frac{N!}{X!(N-X)!}p^X(1-p)^{N-X},
\end{equation}
where $S$ is the total number of counts in the source region, $B$ is the total number of counts in the background region, \mbox{$N=S+B$}, and $p=1/(1+BACKSCAL)$, with $BACKSCAL$ being the ratio of the background and source region areas. $P_B$ returns the probability that a statistical fluctuation of the background produces the observed number of counts $S$, assuming that there is no real source. Therefore, $(1-P_B)$ is the significance of the detection of a source.
We set the detection threshold to $P_B=5\times10^{-4}$, corresponding to a detection significance of 99.95\% (i.e. $\approx3.5\sigma$ in the Gaussian approximation).

For sources observed with \chandra, we performed aperture photometry on the $0.5-2$ and \mbox{$2-7$}~keV images using circular extraction regions centered at the positions of the targets\footnote{The \textit{WISE} all-sky survey astrometric frame is tied to 2MASS, and  \textit{WISE} positions have typical statistical uncertainties of $\approx0.3$ arcsec. See http://wise2.ipac.caltech.edu/docs/release/allsky/expsup/.} and with radii of $R=2$ arcsec, assuming that they contain 100\% of the source counts. Note that this is a good approximation also for W0134--2922, which lies 3.6 arcmin from the aim point of its observation. This choice of the extraction region was driven by considering the cumulative effect of the \chandra\, PSF ($\lesssim1$ arcsec), and \textit{WISE} and \textit{Chandra} positional uncertainties ($\lesssim0.5$ arcsec).
Source photometry for \xmm\, targets has been assessed in circular regions with radii of 15 arcsec, assumed to contain 70\% of the source counts.
Background levels have been assessed in annular regions centered at the positions of the targets, with inner radii sufficiently large to avoid the inclusion of a significant number of source counts, or in nearby regions free from detected sources.

\subsection{Results for targets of the \chandra\, snapshot survey }\label{det_snap}
All targets of the \chandra\, snapshot survey have zero counts in the $0.5-2$~keV band except for  W1248--2154, which has one count.\footnote{W1248--2154 is discussed in detail in \S~4.2.1 of \cite{Tsai15}. A $z=0.339$ galaxy is detected in the $K$ band at a projected distance of $\approx1.3$ arcsec from W1248--2154. The authors investigated a possible lensing effect, but did not find evidence for that. The single photon detected in the $0.5-2$~keV band close to W1248--2154 is at a distance $\approx1.7$ arcsec from the $z=0.339$ galaxy. In any case, if it was emitted by that galaxy, the  average emission from our sample would be even harder than we found (see \S~\ref{HR_sec}), reinforcing our conclusions (see \S~\ref{L_NH_sec} and \S~\ref{conclusions}).} In the \mbox{$2-7$ keV} band, W0410--0913 and W0831+0140 have two counts, W0255+3345 and W2042--3245 have one count each, while the remaining targets have no counts. 
None of the sources is significantly detected in the considered energy bands.

We computed the constraints on the net counts in the $0.5-2$ and $2-7$ keV bands for each individual object by deriving the probability distribution function of net counts with the method of \citet[see their Appendix A3]{Weisskopf07}, which correctly accounts for the Poisson nature of both source and background counts.\footnote{Throughout this paper, whenever we consider the different areas in which source and background photometry are evaluated, we also account for the different average exposure times.} 
Then, we used XSPEC 12.9.0 \citep{Arnaud96} to compute upper limits on net count rate, flux, and intrinsic luminosity for the individual sources, accounting for the specific response matrices.
 
In this procedure, a spectral shape must be assumed to weight the energy-dependent effective area of \textit{Chandra}. We assumed the MYTorus\footnote{http://www.mytorus.com/} model \citep{Murphy09} to describe the X-ray spectral shape of our sources, with an additional Galactic absorber with column density set to the value along the line of sight of each source (spanning a range \mbox{$N_H=1-8\times10^{20}\,\nhunits$}). We fixed the inclination angle to $90$ deg, the photon index to $\Gamma=2$, the intrinsic column density to $N_H=10^{24}\nhunits$,  and the relative normalization between the zeroth order, scattered, and line components to unity. We show in \S~\ref{HR_sec} that this model, and in particular the large column density, is consistent with the average hardness ratio of the snapshot survey targets, derived from a stacking procedure, and with results for individually detected sources.
Tab.~\ref{phot_tab} presents the results for the individual sources. The estimated values of the intrinsic luminosities are sensitive to the assumed spectral model, and, in particular, to the assumed value of intrinsic column density, while count rates and fluxes do not vary significantly with different choices. To partially mitigate this issue, constraints on intrinsic luminosities are derived from the upper limits on the count rate in the hard band, which is less affected by obscuration (and associated uncertainties) than softer energies. Errors on fluxes and luminosity depend only on the net-count uncertainties.

\begin{table*}
	\caption{Constraints on the photometry and spectral properties of individual targets. Uncertainties for detected sources are reported at the $68\%$ confidence level, while for undetected objects we report the upper limits corresponding to the 90\% confidence level. Column density is derived by spectral analysis for W0116--0505 and W0220+0137, from hardness-ratio analysis for W0134--2922 (assuming the nominal value), and set to $N_H=10^{24}\nhunits$ for the remaining, undetected sources. Rest-frame $2-10$~keV, absorption-corrected luminosities are derived from the observed $2-7$ keV band (for \chandra) or $2-10$ keV band (for \xmm), which are less affected by obscuration, and associated uncertainties, than the $0.5-2$~keV band.}\label{phot_tab}
	\begin{tabular}{cccccccc}
		\hline
		\multicolumn{1}{c}{\textit{WISE}} &
		\multicolumn{2}{c}{Net counts}&
		\multicolumn{1}{c}{$N_H$ $[10^{23}\nhunits]$} &
		\multicolumn{2}{c}{Flux [$10^{-15}$ erg cm$^{-2}$ s$^{-1}$]}&
		\multicolumn{2}{c}{$L$ [$10^{45}$ erg s$^{-1}$]} \\
		\multicolumn{1}{c}{} &
		\multicolumn{1}{c}{Soft band} &		
\multicolumn{1}{c}{Hard band$^a$} &
		\multicolumn{1}{c}{} &
\multicolumn{1}{c}{Soft band} &		
\multicolumn{1}{c}{Hard band$^a$} &	
		\multicolumn{1}{c}{$2-10$ keV} &\\				
		\hline
				\multicolumn{7}{c}{Targets of the \chandra\, snapshot survey}\\
		W0134--2607 & $<2.3$ & $<2.3$  &10  &  $<4.1$ & $<16.1$ & $<2.5$ \\
		W0149+2350 & $<2.3$ & $<2.3$  & 10  &  $<3.8$ &  $<14.7$ & $<4.4$\\
		W0255+3345 & $<2.3$ &$<3.8$   &10  &   $<4.0$& $<25.8$& $<5.7$ \\ 
		W0410--0913  & $<2.3$ & $<5.3$ &10  &  $<3.6$ & $<32.3$&$<11.6$ \\ 
		W0533--3401  & $<2.3$ & $<2.3$ &10  & $<3.8$& $<15.0$&$<3.8$ \\ 
		W0615--5716  & $<2.3$ &$<2.3$   &10   &  $<3.7$ &$<14.4$	& $<4.7$ \\ 
		W0831+0140 & $<2.3$   & $<5.3$ &10  &   $<3.7$ & $<31.8$&$<13.1$  \\ 
		W0859+4823 & $<2.3$ & $<2.3$  &10  &  $<3.6$ &$<14.5$ & $<4.4$ \\ 
		W1248--2154 & $<3.9$  &$<2.3$  &10  &  $<6.1$&$<14.5$& $<4.6$\\ 
		W1322--0328 &  $<2.3$ &$<2.3$ &10  &   $<3.8$ & $<14.8$& $<4.0$ \\  
		W2042--3245 & $<2.3$& $<3.9$  &10   &  $<3.7$ &$<23.0$ & $<9.8$\\ 
		W2201+0226  & $<2.3$ & $<2.3$ &10  &   $<3.9$ & $<15.1$ & $<3.8$\\
		W2210--3507 &$<2.3$ & $<2.3$  & 10 &   $<3.8$&$<14.7$ & $<3.5$\\ 
		W2246--7143 & $<2.3$  & $<2.3$ &10  &  $<3.6$ & $<14.1$&$<4.7$\\ 
		W2305--0039 & $<2.3$ & $<2.3$ &10  &  $<3.7$& $<14.3$&$<4.0$\\ 

		\multicolumn{7}{c}{Targets of long-exposure observations}\\		
		W0116--0505           & $15.5^{+4.3}_{-3.7}$ & $56.1^{+7.9}_{-7.2}$ & $12.6^{+2.8}_{-2.5}$&    $1.2^{+0.3}_{-0.3}$ & $17.8^{+4.3}_{-3.8}$ &$3.1^{+0.8}_{-0.6}$\\
		W0134--2922           & $2.9^{+2.1}_{-1.4}$ & $2.7^{+2.1}_{-1.5}$     & $2.6$ &     $0.4^{+0.3}_{-0.2}$ & $1.2^{+1.0}_{-0.7}$&  $0.2^{+0.1}_{-0.1}$\\
		W0220+0137            & $6.5^{+3.0}_{-2.3}$ & $11.2^{+3.8}_{-3.1}$ & $2.8^{+2.9}_{-1.6}$ &    $0.7^{+0.6}_{-0.3}$ & $2.1^{+1.9}_{-0.8}$  &  $0.3^{+0.3}_{-0.1}$\\
		W1838+3429$^{b}$  & $<3.0$ & $<13.3$ &   $10$ &  $<1.0$ & $<25.1$    & $<5.1$\\
		W2246--0526$^{b}$ & $<15.5$ & $<14.8$ &   $10$ & $<2.2$ & $<9.4$   &  $<3.8$\\		         
		\hline
	\end{tabular}\\
$^a$ Hard band is defined as the $2-7$ keV band for \chandra\, and $2-10$ keV band for \xmm. 
$^b$ Fluxes and luminosity are corrected for the 70\% of the \xmm\, PSF included in the source-extraction region.
\end{table*}

\subsection{Results for sources covered by long-exposure observations }\label{results_indiv}

\subsubsection{W0116--0505}
\begin{figure} 
	\centering
	\includegraphics[width=90mm,keepaspectratio]{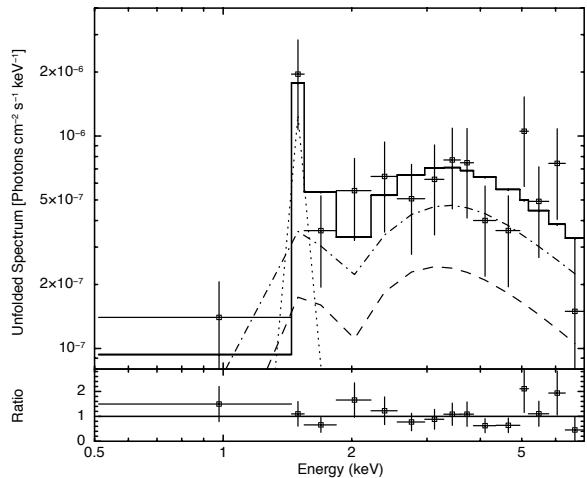}
	\caption{Upper panel: unfolded X-ray spectrum of \mbox{W0116--0505}, rebinned ($2\sigma$ per bin) for display purposes, and total best-fitting model (solid line). The transmitted component (i.e., absorbed power-law), scattered continuum, and iron $K_\alpha$ emission-line component are shown as dashed, dot-dashed, and dotted lines, respectively. Lower panel: ratio between the data points and the model expectations. The best-fitting MYTorus model is consistent with being Compton thick.}
	\label{W0116}
\end{figure}

According to Eq.~\ref{PB}, W0116--0505 is detected in both the \chandra\, soft and hard bands with \mbox{$P_B\ll5\times10^{-4}$} and $\approx70$ net counts in the $0.5-7$ keV band (see Tab.~\ref{phot_tab}). We performed a basic spectral analysis using the MYTorus model in XSPEC and using the Cash statistics \citep{Cash79}. Due to the limited photon counting statistics, we fixed the photon index to $\Gamma=2$. 
For the same reason, we left free the normalization of the scattered and line components with respect to the primary power-law during the fit and when computing the error on column density, but we froze it when computing the errors on flux and luminosity. Best-fitting parameters are reported in Tab.~\ref{phot_tab}. The best-fitting model is shown in Fig.~\ref{W0116}. In particular, the column density is constrained in the range $(1.0-1.5)\times10^{24}\nhunits$, making W0116--0505 a Compton-thick QSO candidate. We computed the rest-frame iron $K\alpha$ equivalent width ($EW$) following the procedure described in the MYTorus manual, where caveats are also discussed, and found $EW=0.9\pm0.2$ keV, consistent with the expectations for Compton-thick AGNs. Its intrinsic X-ray luminosity is the highest among Hot DOGs currently detected in X-ray observations.

\subsubsection{W0134--2922}
We considered only the (serendipitous) \chandra\, observation (see \S~\ref{sample_indiv}), in which the source is detected at an off-axis angle of $\approx3.6$ arcmin. W0134--2922 is detected in the soft band with a significance of $(1-P_B)=99.985\%$ (corresponding to a $\approx3.8\sigma$ detection in the Gaussian approximation) and undetected in the hard band ($1-P_B=99.654\%$, corresponding to $\approx2.9\sigma$).

The small number of total detected counts in the source region (3 in both the soft and hard bands) prevented us from performing a spectral fitting analysis. 
The hardness ratio, defined as
\begin{equation}
HR=\frac{H-S}{H+S},
\end{equation}
where $S$ and $H$ are the net (i.e., background-subtracted) counts in the soft and hard bands, respectively, depends on the observed-frame spectral shape of a source, and therefore provides basic estimates of parameters such as photon index and column density. We used the BEHR\footnote{http://hea-www.harvard.edu/astrostat/behr/} code \citep{Park06}, which accounts for the Poisson nature of both source and background counts, and derived $HR=-0.03\pm0.37$. Assuming the MYTorus model with photon index fixed to $\Gamma=2$ and accounting for the appropriate response matrices, the nominal value of the hardness ratio corresponds to an obscuring column density of $N_H=2.6\times10^{23}\nhunits$. Although considering the uncertainties on $HR$ we can only set an upper limit on the column density ($N_H<6.7\times10^{23}\,\nhunits$),
we estimated the flux and intrinsic luminosity fixing it to the nominal value. This procedure is similar to what we did in \S~\ref{det_snap}, but, instead of assuming $N_H=10^{24}\,\nhunits$, we make use of the information derived from the hardness ratio. Results are reported in Tab.~\ref{phot_tab}.

\subsubsection{W0220+0137}
W0220+0137 is detected with significance \mbox{$P_B\ll5\times10^{-4}$} in both the \chandra\, soft and hard bands.
For this source, we repeated the same analysis performed for W0116--0505, but, due to the smaller number of detected counts, we fixed the normalization of the scattered component with respect to the intrinsic power-law to unity during the fit. Best-fitting parameters are reported in Tab.~\ref{phot_tab}.

\subsubsection{W1838+3429}
This source is not detected in any of the \xmm\, soft ($0.5-2$ keV), hard ($2-10$ keV), or full ($0.5-10$ keV) bands by any of the individual \xmm\, instruments.
For simplicity, we consider only the PN camera, which has the highest sensitivity.
We computed upper-limits at the 90\% confidence level on its photometry (Tab.~\ref{phot_tab}), assuming its spectrum is well described by the MYTorus model with $N_H=10^{24}\nhunits$, similarly to what we did in \S~\ref{det_snap}. We note, however, that the particular choice of intrinsic $N_H$ does not strongly affect the derived upper limit on luminosity because $L_X$ is derived from the flux in the observed-frame hard band, which is marginally absorbed even by large column densities at high redshift.

\subsubsection{W2246--0526}
This Hot DOG is the most luminous galaxy in the sample, and one of the most-luminous known galaxies in the Universe.\footnote{To our knowledge, this galaxy has the highest bolometric luminosity derived from SED fitting. Bolometric luminosities for other galaxies with similar or slightly higher power have been derived via scaling relations with optical emission-line luminosities. We also note that conservative estimates of bolometric luminosities were derived by \citet[see their section 3.3]{Tsai15}, and thus may be underestimated.} The source is not significantly detected in either of the \xmm\, bands by any of its instruments, although a somewhat enhanced signal is perhaps visible in the PN camera in the full band ($1-P_B\approx98.76\%$, corresponding to $\approx2.5\sigma$ in the Gaussian approximation). 
We also tried performing the standard \xmm\, detection procedure on the data from the combined cameras. Using EMLDETECT,\footnote{https://heasarc.gsfc.nasa.gov/docs/xmm/sas/help/emldetect/node3.html} W2246--0526 is not detected with a DET\_ML = 1.67, which is equivalent to a false-detection probability of 0.19.
We considered only the PN dataset and derived upper limits at the 90\% confidence level on the source counts, fluxes and luminosity, assuming the same spectral model used for W1838+3429.

\section{Stacking procedure and photometry of snapshot targets}\label{stack_sec}
Because of the high non-detection rate of the Hot DOGs targeted by our \chandra\, snapshot survey, we performed a stacking analysis to derive their average X-ray count rate.
The stacking procedure is justified by the homogeneous X-ray observational properties of this sample (i.e, they are observed with the same instrument for the same exposure time), which is also fairly uniform in terms of redshifts and bolometric luminosities.
For each of the 15 observations we created an image in a given energy range, centered at the position of the target. The individual images were then summed with the \textit{dmimgcalc} tool. 
Similarly, we summed the exposure maps of each observation.
We also extracted response matrices and ancillary files with the \textit{specextract} tool and averaged them with the \textit{addrmf} and \textit{addarf} scripts of the HEASOFT 6.19 software package.\footnote{https://heasarc.gsfc.nasa.gov/lheasoft/}

\begin{figure*} 
	\centering
	\includegraphics[width=180mm,keepaspectratio]{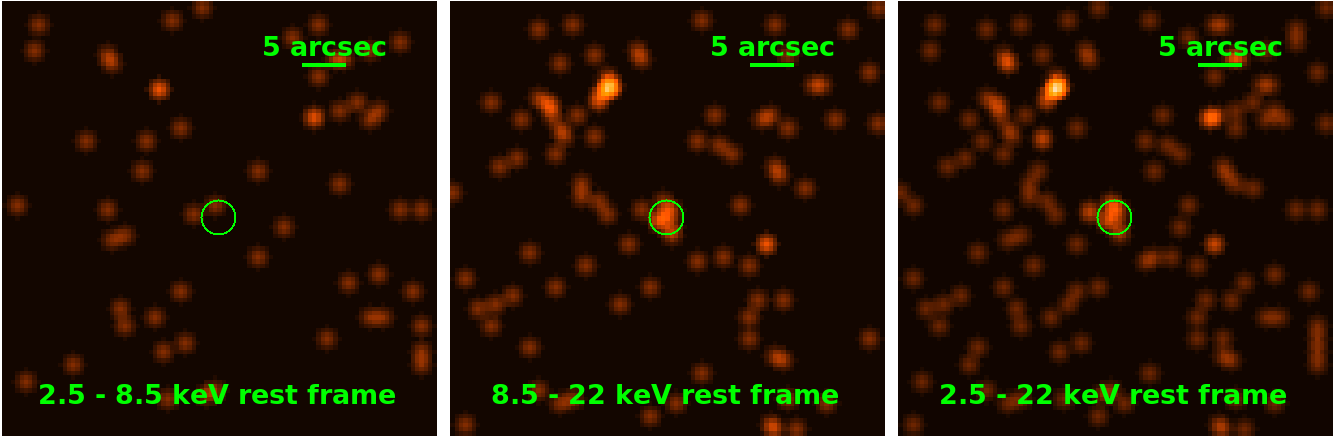}
	\caption{From left to right: stacked and smoothed X-ray images in the rest-frame $2.5-8.5$, $8.5-22.0$, and $2.5-22.0$ keV bands. Each panel has a dimension of $101\times101$ pixels (about $50\times50$ arcsec). Solid circles have radii of 4 pixels ($\approx2$ arcsec) and represent the extraction regions we used to compute the photometry. The significant emission in the upper central parts of the images (especially in the hard and full bands) is due to a serendipitously detected X-ray source (with 7--9 counts) in the proximity ($\approx16$ arcsec offset) of W2246--7143.}
	\label{stack}
\end{figure*}

Fig.~\ref{stack} presents the stacked images in the rest-frame $2.5-8.5$, $8.5-22.0$, and $2.5-22.0$ keV bands (from left to right). The $2.5-22.0$ keV band is the widest common rest-frame energy interval probed by \chandra\, for this sample. The  $2.5-8.5$~keV and $8.5-22.0$~keV bands were chosen to approximately match the more standard observed-frame $0.5-2$ and $2-7$ keV bands. Similarly to what we did in \S~\ref{detection}, we computed the photometry in each band using circular extraction regions with radii of 4 pixels ($\approx2$ arcsec, solid circles in Fig.~\ref{stack}). 
The extension of a serendipitously detected X-ray source (with 7--9 total counts) close to W2246--7143 and visible in the upper central areas of the stacked images can be compared with our extraction regions. 

Our source regions encompass 1, 5, and 6 total counts in the $2.5-8.5$, $8.5-22.0$, and $2.5-22.0$ keV bands, respectively. Each single object contributes $\leq2$ counts to the stacked photometry, which is therefore not dominated by any individual source. In particular, only W0410--0913 contributes two counts in the $2.5-22.0$~keV band, while all the other sources contribute one or zero counts.\footnote{We note that W0831+0140 was reported with two counts in \S~\ref{det_snap} in the observed-frame $2-10$ keV band, while here we are considering the rest-frame $8.5-22$ keV band (i.e., observed-frame $1.7-4.5$ keV for that galaxy).} We stress that the energy bands used in this section are different than those used in \S~\ref{detection} to assess the detection significance of the individual sources. Background levels have been assessed from annular regions with inner and outer radii of 6 and 50 pixels, respectively, resulting in 0.0058, 0.0090, and 0.0148 background counts per pixel in the $2.5-8.5$, $8.5-22.0$, and $2.5-22.0$ keV bands, respectively. We masked the serendipitously detected source (see above in this section) for computing the background photometry. The stacked exposure time is \mbox{$\approx46$} ks, resulting in average net-count rates of \mbox{$<0.5\times10^{-4}$}, \mbox{$1.0^{+0.6}_{-0.4}\times10^{-4}$}, and \mbox{$1.1^{+0.6}_{-0.5}\times10^{-4}$~cts~s$^{-1}$}, in the $2.5-8.5$, $8.5-22.0$, and $2.5-22.0$ keV bands, respectively.

We detect stacked signals in the $8.5-22.0$ and $2.5-22.0$ keV bands, with $P_B=0.9\times10^{-4}$ and $1.3\times10^{-4}$, respectively, corresponding to significances of $\approx99.99\%$ ($\approx3.9\sigma$) in both bands.
 The fraction of detected photons for the subsamples of sources classified as Ly$\alpha$ emitters or presenting AGN features in their optical spectra (see column 6 of Tab.~\ref{obs_tab}, Eisenhardt et al., in prep.) is roughly proportional to the subsample sizes. Therefore, there is no strong evidence for a dependence of X-ray emission on the optical spectral type.

\subsection{Hardness-ratio analysis}\label{HR_sec}
We computed the lower limits on the average $HR$ of the stacked sources at the 68\% and 95\% confidence levels using the stacked photometry. The resulting values are $HR>0.36$ and \mbox{$>-0.08$}, respectively, corresponding to effective photon indexes (i.e., the slope of an unabsorbed power-law) of $\Gamma_{eff}<0.26$ and $<0.40$, respectively. 
These values strongly suggest our sources are heavily obscured.

\begin{figure} 
	\centering
	\includegraphics[width=85mm,keepaspectratio]{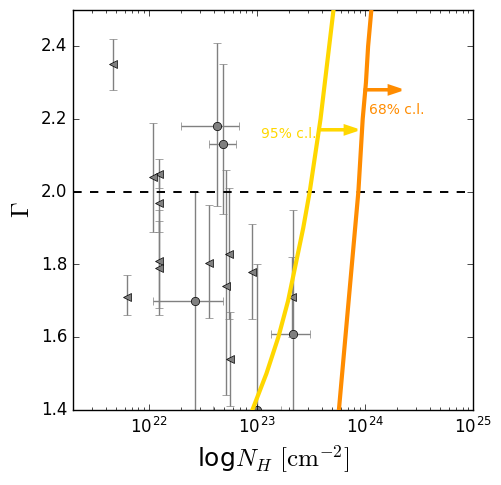}
	\caption{Confidence limits (at 68\% and 95\% levels, in orange and gold, respectively) in the $\Gamma-N_H$ plane obtained from hardness-ratio analysis of the stacked emission of our sample. Grey symbols are from the sample of optical Type 1 QSOs from \citet{Martocchia17}. Leftward-pointing symbols are upper limits (90\% confidence level) on column density. The dashed horizontal line represents the value for $\Gamma$ we assumed in \S~\ref{L_NH_sec}. }
	\label{HR}
\end{figure}

We used the MYTorus model to describe the average X-ray spectral shape of our sources. We fixed the inclination angle to $90$ deg, the relative normalization between the zeroth order, scattered, and line components to unity, and the redshift to $z=3.23$, the median redshift of the stacked sample. We accounted for a Galactic absorber with column density of \mbox{$N_H=4\times10^{20}\,\nhunits$}, the median value along the lines of sight of these targets. 
Then, we ran a set of simulations with the \textit{fakeit} command, using the averaged response matrix and ancillary file, allowing the photon index and the intrinsic column density to vary in the ranges of $\Gamma=1.4-2.5$ and log$N_H[\nhunits]=22-25$. For each pair of these parameters we obtain the expected number of counts we would detect in our soft and hard bands, and thus the expected $HR$. Finally, comparing the expected and measured hardness ratios, we could identify confidence regions for $\Gamma$ and $N_H$, which are presented in Fig.~\ref{HR}. 
According to our $HR$ analysis, the average X-ray spectrum of our targets is obscured by column densities of $N_H\gtrsim8\times10^{23}\,\nhunits$ ($N_H\gtrsim3\times10^{23}\,\nhunits$ ) at 68\% (95\%) confidence level for any reasonable value of the intrinsic photon index ($\Gamma\gtrsim1.7$). 
This column density is significantly larger than the typical values found for optically bright QSOs with similar bolometric luminosities for which X-ray spectral analysis has been performed \citep[e.g.][]{Martocchia17}. Lower levels of obscuration would require very flat and possibly implausible photon indexes.

\section{Results}
\subsection{The $L_X$--$N_H$ plane}\label{L_NH_sec}

We derived joint constraints on the average luminosity and column density of the stacked sources, assuming an intrinsic photon index of $\Gamma=2$ for the primary power-law emission. This is justified by the expected high accretion rate, close to the Eddington limit, onto the SMBHs in Hot DOGs \citep{Wu17}, which results in steep X-ray spectra, following the known relation between these two physical quantities \citep[e.g.,][]{Shemmer08,Fanali13,Brightman13,Brightman16}. Also, the SDSS QSOs with similar luminosities to our sources studied by \cite{Just07} have $\Gamma\approx2$. We make use only of the photometry derived from the stacked image in the full band. Consistent results are derived using the hard band. Using the soft band only, where we did not detect a signal, would not provide useful constraints. As we did for the individual sources, net counts and uncertainties are determined using the method of \cite{Weisskopf07}, which correctly accounts for the Poisson nature of both source and background counts.

\begin{figure*} 
	\centering
	\includegraphics[width=160mm,keepaspectratio]{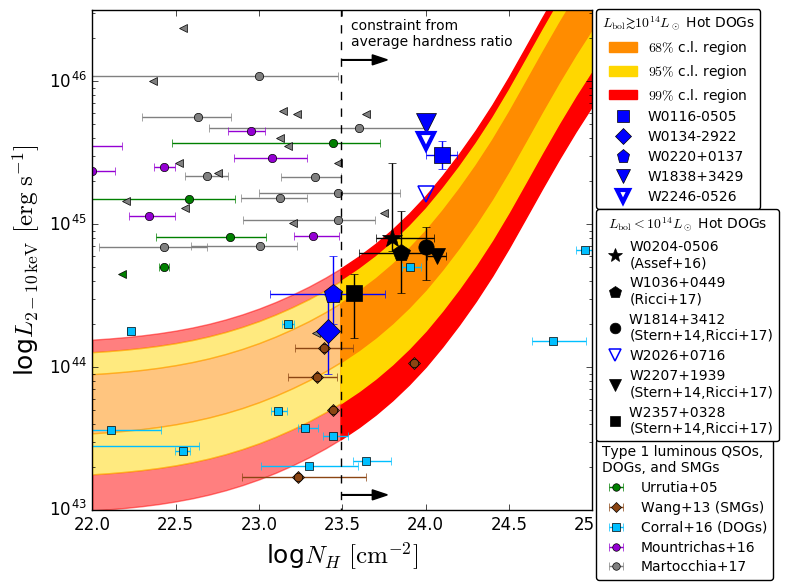}
	\caption{Locations of the Hot DOGs covered by long-exposure X-ray observations in the $L_X$-$N_H$ plane. Blue symbols are from this work, while black symbols are collected from the literature. See Appendix~\ref{appendix} for the analysis on W2026+0716. We also show the joint constraints on the average intrinsic luminosity and column density derived in \S~\ref{L_NH_sec} for the targets of the  \chandra\, snapshot survey as orange, gold, and red stripes, corresponding to the 68\%, 95\%, and 99\% confidence regions, respectively. The vertical dashed line marks the lower limit (at 95\% confidence level) derived from hardness-ratio analysis of the stacked signal (Fig.~\ref{HR}), assuming an intrinsic $\Gamma=2$. For comparison purposes, we include the luminous Type~1 QSOs with bolometric luminosities similar to Hot DOGs from \citet[green symbols]{Urrutia05}, \citet[grey symbols]{Martocchia17}, and  \citet[purple symbols]{Mountrichas17}. We also show the locations of luminous DOGs from  \citet[cyan symbols]{Corral16} and sub-millimiter galaxies \citep[SMGs;][brown symbols]{Wang13}. Leftward-pointing and downward-pointing triangles are upper limits on $N_H$ and $L_X$, respectively. When possible, error bars and upper limits from the literature have been homogenized to 68\% and 90\% confidence levels, respectively, assuming a Gaussian approximation.
	}
	\label{L_NH}
\end{figure*}

We computed with XSPEC the intrinsic luminosity required to match the observed count rate (and associated uncertainties), considering a grid of column density log$N_H[\nhunits]=22-25$. The stacked count rate is assumed to be representative of the average X-ray emission of the stacked Hot DOGs in the rest-frame 2.5--22.0 keV band at their median redshift ($z=3.23$).
We used the MYTorus model with the same configuration used in \S~\ref{HR_sec} and the average response matrix and ancillary file.
We could thereby retrieve confidence regions in the $L_X$--$N_H$ plane, presented in Fig.~\ref{L_NH}. We stress that, being derived from stacking analysis, these are the confidence regions of the \textit{average} $L_X$ at a given $N_H$ for the stacked sources. We also mark with a vertical line the lower limit (at 95\% confidence) of the column density derived from the hardness ratio analysis, assuming $\Gamma=2$.

The assumed spectral model describes the observed emission in the case of a toroidal distribution of obscuring material. Although this model is the best physically motivated model available, it may not be a correct representation for our sources, which are extremely luminous and possibly in a post-merger stage. Moreover, the obscuring material may be distributed on larger scales than a compact torus. We repeated our analysis using a more simple absorbed power-law model, widely used in the literature \cite[e.g., see][for a recent example involving luminous QSOs]{Martocchia17} and found consistent results within the uncertainties. Similarly, varying the fixed parameters of our models (i.e., allowing the torus inclination angle and the photon index to vary in the ranges $\Theta\approx70-90$~deg and $\Gamma=1.8-2.2$, respectively) does not affect significantly the results. Low inclination angles ($\Theta\lesssim70$ deg) would result in an increasing probability of our sources being intrinsically weak and unobscured. This is due to the geometry of the model, which assumes an opening angle of $60$ deg for the torus. Therefore, at comparable or lower inclination angles the line of sight intercepts at most marginally the obscuring material. However, this is not consistent with the constraints on column density we set from the hardness ratio analysis (\S~\ref{HR_sec}).

In Fig.~\ref{L_NH} we also include the individual Hot DOGs whose X-ray emission has been studied in this work (blue symbols, \S~\ref{results_indiv} and Appendix~\ref{appendix}) or collected from the literature (black symbols, \citealt{Stern14, Assef16, RicciC17}). Their positions on the plane agrees well with the constraints from the stacking analysis. In particular, all of them are obscured by intrinsic column densities of log$N_H[\nhunits]\gtrsim23.5$, consistent with the hardness-ratio analysis of the stacked sources. This result strongly suggests that Hot DOGs are typically obscured in X-rays by column densities close to or even exceeding the Compton-thick limit.

We compare the locations of Hot DOGs in the $L_X$--$N_H$ plane with other extremely luminous QSOs. In particular, we consider the red broad-line QSO sample from \cite{Urrutia05},\footnote{We do not include one of their sources, namely FTM 0134--0931, as its luminosity is reported by the authors to be enhanced by gravitational lensing.} and the luminous Type 1 WISE/SDSS selected QSO samples from \cite{Martocchia17} and \cite{Mountrichas17}. Fig.~\ref{L_NH} shows that Hot DOGs occupy a different region of the plane than extremely luminous, Type 1 QSOs with brighter optical magnitudes. In particular, the latter are typically much less obscured and have comparable \mbox{X-ray} luminosities. In fact, most of the Hot DOGs covered by long X-ray exposure have log$L_X\,[\lunit]\gtrsim45$, similar to broad-line QSOs. Such a high average luminosity is also suggested by our constraints derived from the hardness ratio of the stacked signal. On the other hand, a few of the Hot DOGs detected in deeper
	exposures imply slightly lower luminosities.

We also include the DOG sample of \cite{Corral16} and the sub-millimeter galaxies studied by \cite{Wang13}. These objects span a wide range of column density, with some DOGs, in particular, obscured by Compton-thick material, and have lower X-ray luminosities\footnote{This statement is also reinforced by the fact that only X-ray detected objects are considered, while X-ray information for undetected sources are not reported by those works.} than Hot DOGs, suggesting that they are accreting at a lower rate. Finally, we note that some SDSS-selected Type-2 AGN have been found to have log$L_X\,[\lunit]\approx44$ and log$N_H\,[\nhunits]\approx24$ \citep[e.g.,][]{Lansbury15}, close to the location of Hot DOGs in this plane.

\subsection{The relations between X-ray, infrared, and bolometric luminosity}

X-ray luminosity and mid-IR  emission (usually traced by the luminosity at $6\,\mu$m; i.e., $L_{6\mu\mathrm{m}}$) are known to correlate almost linearly in low-to-moderate luminosity AGNs \citep[e.g.;][]{Lutz04,Fiore09,Gandhi09, Lanzuisi09}. However, such a correlation has been recently found to flatten at high luminosity \citep[e.g.][]{Stern15,Chen17}. Moreover, a luminosity-dependent relation between X-ray and bolometric luminosity is in place for accreting SMBH \citep[e.g.][]{Marconi04, Lusso12}, but it is still not well assessed for the extreme tail of the QSO population.
It is therefore worth investigating these two relations for our sample of Hot DOGs, which host some of the most-luminous QSOs in the Universe caught during a key evolutionary stage characterized by heavy obscuration.

\begin{figure*} 
	\centering
	\includegraphics[width=160mm,keepaspectratio]{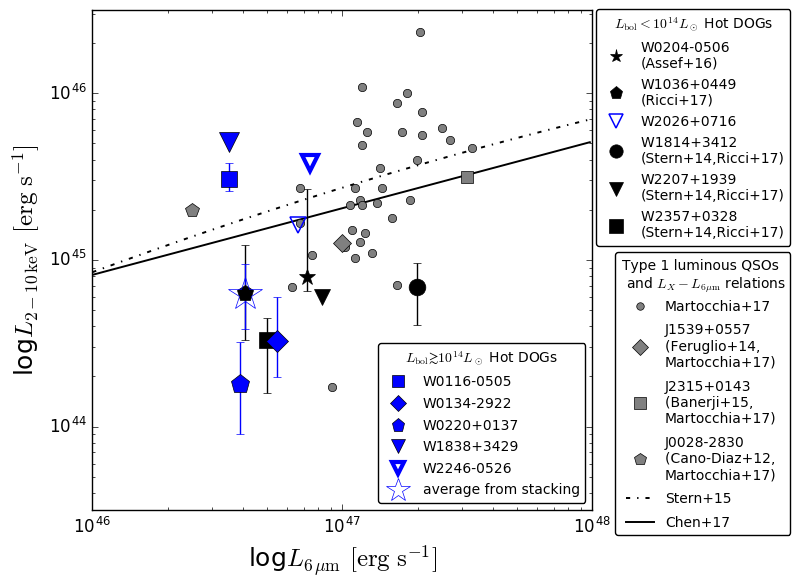}
	\caption{Intrinsic X-ray luminosity versus mid-IR luminosity for Hot DOGs. Blue and black symbols represent individual sources presented in this work or collected from the literature. See Appendix~\ref{appendix} for the analysis on W2026+0716. Error bars are given at 68\% confidence level. Downward pointing arrows represent upper limits on the X-ray luminosities at 90\% confidence level. The average X-ray luminosity for stacked sources (derived from Fig.~\ref{L_NH} assuming a column density of $N_H=10^{24}\,\nhunits$) is marked at their median $L_{6\mu\mathrm{m}}$ with a blue star. We also add for comparison the sample of luminous optical Type 1 QSOs collected by \citet{Martocchia17}, and we plot the relations of \citet{Stern15} and \citet{Chen17}. }
	
	\label{LX_L6mu}
\end{figure*}

Fig.~\ref{LX_L6mu} presents the X-ray luminosities of Hot DOGs as a function of $L_{6\mu\mathrm{m}}$, compared with optical Type 1 QSOs with similar luminosities. 
We also plot the average X-ray luminosity derived from the stacking analysis of \chandra\, snapshot targets, assuming a column density of $N_H=10^{24}\,\nhunits$, at the position of their median $L_{6\mu\mathrm{m}}$.
Hot DOGs fall on average slightly below the \cite{Stern15} and \cite{Chen17} relations. 

We warn that $L_{6\mu m}$ has been computed under the assumption of isotropic emission \citep{Tsai15}, while a degree of anisotropy is expected, especially for heavily obscured sources \citep[e.g.][]{Pier92,Fritz06, Nenkova08a,Nenkova08b, Stalevski12,Mateos15}, which would lead us to underestimate the infrared luminosity. Also, the average hardness ratio of stacked sources is consistent with column densities as low as log$N_H\,[\nhunits]\approx23.5$. In these cases, the average X-ray luminosity of Hot DOGs would be in even stronger disagreement with those relations, which are derived for Type 1 QSOs.

\begin{figure*} 
	\centering
	\includegraphics[width=160mm,keepaspectratio]{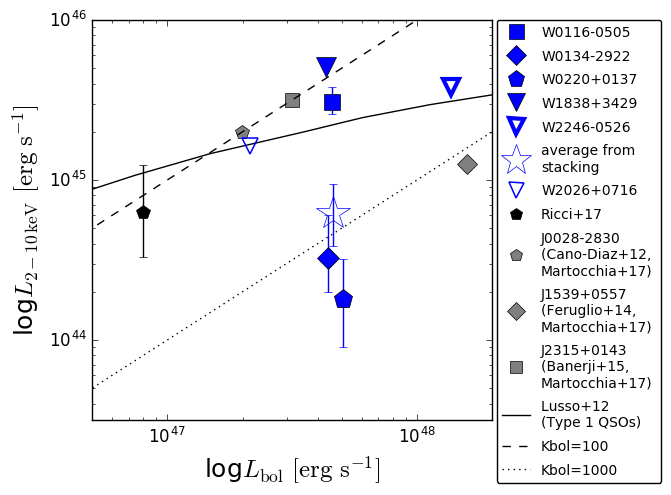}
	\caption{Intrinsic X-ray luminosity versus bolometric luminosity for Hot DOGs. Blue and black symbols represent individual sources presented in this work or collected from the literature, if $L_{\mathrm{bol}}$ is reported in the original paper. See Appendix~\ref{appendix} for the analysis on W2026+0716. Error bars are given at 68\% confidence level. Downward pointing arrows represent upper limits on the X-ray luminosities at 90\% confidence level. The average X-ray luminosity for the stacked sources (derived from Fig.~\ref{L_NH} assuming a column density of $N_H=10^{24}\,\nhunits$) is marked at their median $L_{\mathrm{bol}}$ with a blue star. We also add for comparison three luminous optical Type 1 QSOs collected by \citet{Martocchia17} and originally presented by \citet{Cano-Diaz12}, \citet{Feruglio14}, and \citet{Banerji15}. The luminosity-dependent relation between $L_X$ and $L_{\mathrm{bol}}$ of \citet{Lusso12} is reported for reference, as well as the expected $L_X$ assuming constant bolometric corrections of $K_{\mathrm{bol}}=100$ and 1000.
	}
	\label{LX_Lbol}
\end{figure*}

Fig.~\ref{LX_Lbol} presents the X-ray versus bolometric luminosities of Hot DOGs and Type 1 QSOs, for those whose  bolometric luminosities have been included in the original papers. We mark in the figure the expected intrinsic \mbox{X-ray} luminosity derived applying the luminosity-dependent bolometric correction ($K_{\mathrm{bol}}$) of \cite{Lusso12}, and constant corrections $K_{\mathrm{bol}}=100$ and 1000.
Although the small number of sources prevents us from drawing strong conclusions, we note that Hot DOGs appear to span a wide range of bolometric corrections, but on average are located close to the $K_{\mathrm{bol}}=1000$ curve, significantly below the relation derived by \cite{Lusso12} extrapolated to these extreme luminosities (with the notable exception of W0116--0505). We also note that the bolometric luminosities for our sources have been derived with a conservative approach by \citet[see their section 3.3]{Tsai15}.
An average column density of $N_H\gtrsim2-3\times10^{24}\,\nhunits$ would be required in order to match the expected $L_X-L_{6\mu\mathrm{m}}$ and $L_X-L_{\mathrm{bol}}$ relations. Larger samples of luminous obscured QSOs covered by X-ray observations are needed to probe this luminosity regime.

\section{Discussion}\label{conclusions}
\subsection{Insights into the nature of Hot DOGs}

In Fig.~\ref{L_NH}, we compare the column densities derived through spectral or hardness-ratio analysis of Hot DOGs
 with the results derived by \cite{Urrutia05}, \cite{Martocchia17}, and \cite{Mountrichas17} for optical Type 1 QSOs. \cite{Martocchia17} investigated the X-ray properties of a sample of \textit{WISE}/SDSS detected Type 1 QSOs with \mbox{$L_{bol}\gtrsim2\times10^{47}\,\lunit$} at $z=2-4$. According to the basic evolutionary path described in \S~\ref{intro}, these objects are caught in a phase when most of the obscuring material is being swept away, possibly by powerful radiatively driven outflows, allowing for the direct detection of radiation from the active nuclei, and broad emission lines in particular. \cite{Mountrichas17} cross-correlated a sample of log$L_{6\mu\mathrm{m}}\mathrm{[erg\,s^{-1}]}>46.2$ WISE/SDSS selected QSOs with the 3XMM X-ray catalog. 
  Qualitatively similar to the \cite{Martocchia17} and  \cite{Mountrichas17} sample are the optical Type~1, luminous red QSOs studied by \cite{Urrutia05} at $z=0.4-2.2$, which occupy a similar region of the $\Gamma$--$N_H$ plane, and the dust-reddened, optically selected Type~1 QSO ULASJ1234+0907 at $z=2.5$, whose X-ray spectrum was analyzed by \cite{Banerji14} with best-fitting values $\Gamma=1.68^{+0.25}_{-0.22}$ and $N_H=9.0^{+1.0}_{-0.8}\times10^{21}\,\nhunits$.

Our sources, selected on the basis of \textit{WISE} colours, plausibly are in an earlier evolutionary stage and on average more obscured, as can be inferred from Fig.~\ref{HR}.  In these objects, radiative feedback, if already triggered (see, e.g., \citealt{Diaz-Santos16} for evidence of outflowing gas in W2246--0526), has not yet been able to sweep away the obscuring material. The X-ray detected, Type 2 DOG at $z=1.59$ studied by \cite{Brusa10}, where it is referred to as COSMOS XID 2028 ($N_H\approx10^{22}\nhunits$, $L_X\approx10^{45}\lunit$), is probably caught in the transition phase between the buried QSO and the optical Type~1 QSO stages. Indeed, extensive optical-to-mm follow-up campaigns revealed strong signatures of molecular and ionized outflows from its nuclear region \citep{Brusa15,Brusa15b,Brusa16}.

In addition to the Hot DOGs whose X-ray emission has been analyzed in this work (\S~\ref{results_indiv}), we also report results collected from the literature for Hot DOGs with lower bolometric luminosities.
\cite{Assef16} analyzed the X-ray spectrum of W0204--0506,\footnote{This object is among the subpopulation of Hot DOGs showing enhanced rest-fram UV emission, see \cite{Assef15} for a discussion on its origin.} serendipitously observed with \chandra\, for 160 ks. As for our targets, this galaxy was selected as a Hot DOG, and indeed is consistent with our $\Gamma-N_H$ constraints. It has a hard X-ray spectrum and a tentatively detected iron~$K_\alpha$ line, both strongly suggestive of heavy obscuration.  A 42 ks \textit{XMM-Newton} spectrum of the Hot DOG W1835+4355 was investigated by \cite{Piconcelli15}. They reported that it is consistent with a reflection-dominated spectral model, due to absorption by material with log$N_H\gtrsim10^{24}\nhunits$. Compton-thick obscuration was also proposed by \cite{Stern14} to explain the non-detections or faint-detections of three Hot DOGs observed with $20-30$ ks observations with \mbox{\textit{NuSTAR}} and \textit{XMM-Newton}. 
Similar to these W1W2-selected Hot DOGs is W1036+0449, a galaxy selected by \cite{RicciC17} with a modified technique aimed at identifying Hot-DOG analogs at $z\approx1$. Exploiting \textit{NuSTAR} and \textit{Swift} coverage of this object, \cite{RicciC17} performed a spectral analysis and reported large values of obscuring column density ($N_H=7.1^{+8.1}_{-5.1}\times10^{23}\,\nhunits$) and luminosity (log$L_X[\lunit]=44.80^{+0.29}_{-0.28}$).
Instead of focusing on a few targets, our \chandra\, snapshot strategy allowed us to study economically a sample of a statistically significant size of the most-luminous Hot DOGs. With the addition of the five sources covered by longer exposures, we were able to present the X-ray coverage of all the $L_{\mathrm{bol}}\gtrsim10^{14}\,L_{\odot}$ Hot DOGs currently known.

The allowed regions presented in Fig.~\ref{HR}  and Fig.~\ref{L_NH} for stacked sources, together with the results for individual objects, strongly suggest that Hot DOGs are heavily obscured, with column density close to or even exceeding the Compton-thick limit and intrinsic luminosity log$L_X\,\mathrm{[\lunit]}=44-45.5$. It is worth noting that larger column densities than those derived for Type 1 QSOs with similar luminosities are expected from the evolutionary framework in which Hot DOGs host deeply buried, rapidly accreting SMBHs, that eventually will produce powerful radiation-driven outflows and shine as blue Type 1 luminous QSOs.

\subsection{Future prospects}
In this section, we focus on the targets of our \chandra\, snapshot survey which allowed us to gather a statistically significant sample of Hot DOGs with economical X-ray observations. Through the stacking analysis described in \S~\ref{stack_sec}, we can place constraints on the \chandra\, and \xmm\, exposure times needed (on average) to detect enough photons to perform a basic spectral analysis of typical Hot DOGs.
Assuming a heavily obscured ($N_H=10^{24}\nhunits$) power-law with $\Gamma=2$ at $z=3.23$ (the median redshift of our stacked sample), the \textit{average} \chandra\, count rate of our sources in the rest-frame $2.5-22$~keV band ($0.11^{+0.06}_{-0.05}$~counts~ks$^{-1}$ translates\footnote{We used WebPIMMS, https://heasarc.gsfc.nasa.gov/cgi-bin/Tools/w3pimms/w3pimms.pl. We note that WebPIMMS does not account for a possible reflection component.} into average count rates of $\approx0.07-0.19$ and $\approx0.33-94$~cts~ks$^{-1}$ for \chandra\, (observed-frame \mbox{$0.5-7$ keV} band) and \xmm\, (observed-frame \mbox{$0.5-10$ keV} band), respectively. 

Therefore, expensive $\approx500-1400$~ks and \mbox{$\approx400-1200$~ks} observations of Hot DOGs with \chandra\, and \xmm\, would be required on average to gather $\approx100$ and $\approx400$ counts, respectively, in order to perform a reliable spectral analysis. The reported ranges are in agreement with the count rates of the individual Hot DOGs shown in  Fig.~\ref{L_NH}, except for W0116--0505, W0204--0506, and W0220+0137, which are in fact slightly above the 68\% confidence region of the intrinsic luminosity.
 \textit{Athena} \citep{Nandra13}, the next-generation X-ray observatory which will be launched in $\sim2029$, will provide the high sensitivity and throughput needed to perform detailed X-ray spectral studies of statistically significant samples of Hot DOGs with relatively economical observations (i.e., a factor of $\approx20$ shorter exposures than \chandra\, to detect the same number of counts).\footnote{We used the \textit{Athena} response files (version 20150326) available at \mbox{http://www.mpe.mpg.de/ATHENA-WFI/public/resources/responses/} under the hypotheses of a mirror module radius $R_{max}=1190$ mm, 2.3 mm rib spacing, and an on-axis target.}

\section{Summary}
We  have presented the \chandra\, and \xmm\, observations of the 20 most-luminous ($L_{\mathrm{bol}}\gtrsim10^{14}L_\odot$) galaxies discovered by \textit{WISE}. We analyzed the X-ray data of 5 sources covered by long exposures, among which is W2246--0526, one of the very most luminous galaxies known. We performed a stacking analysis of the remaining 15 Hot DOGs targeted by a \chandra\, snapshot survey, in order to derive their average properties. We also collected literature results for another 5 slightly less luminous Hot DOGs, and we compare with luminous optical Type 1 QSOs, DOGs, and SMGs.
Our main conclusions are the following:

\begin{enumerate}
\item According to our stacking analysis, Hot DOGs are on average heavily obscured, with \mbox{log$N_H\,\mathrm{[\nhunits]}\gtrsim24$} and $>23.5$ at 68\% and 95\% confidence level, respectively. Basic spectral  or hardness-ratio analysis for sources covered by long-exposure X-ray observations are consistent with these limits. In particular, W0116--0505 is a Compton-thick QSO candidate with best-fitting $N_H=1.0-1.5\times10^{24}\nhunits$.

\item By gathering a statistically significant sample of Hot DOGs, we could show that they occupy a distinct region of the $L_X-N_H$ plane from optical Type~1 QSOs with similar bolometric luminosities. In particular, Hot DOGs are significantly more obscured and have comparable or slightly lower intrinsic X-ray luminosities. 
This finding fits well in an evolutionary scenario in which Hot DOGs are caught during a phase of extremely fast and deeply buried SMBH accretion, possibly subsequent to merger events. Radiative feedback from the nuclear regions then sweeps away the accreting and obscuring material, allowing the detection of luminous blue QSOs. In this context, optical Type 1, dust-reddened QSOs represent the transition phase. Comparing to DOGs and SMGs, some of which are found to be heavily obscured, Hot DOGs are significantly more luminous.

\item Although the relatively small size of the sample prevents us from drawing strong conclusions, we found hints that Hot DOGs may be less X-ray luminous than the expectations from their mid-IR and bolometric luminosities. In particular, Hot DOGs fall on average slightly below the $L_X-L_{6\mu\mathrm{m}}$ relations of \cite{Stern15} and \cite{Chen17}. Notably, these relations were derived for luminous Type 1 QSOs, and they predict lower X-ray luminosities than extrapolations of literature results derived for AGNs with lower luminosities.
Hot DOGs are also, on average, characterized by a higher bolometric correction ($K_{\mathrm{bol}}\approx1000$; defined as the ratio between bolometric and X-ray luminosity) than the expectation from known relations, derived for AGNs with lower luminosities \citep[e.g.,][]{Lusso12}. Only obscuration due to column densities $N_H\gtrsim2-3\times10^{24}\nhunits$ would result in more standard bolometric corrections.

\item We computed predictions of the average \chandra\, and \xmm\, count rates of Hot DOGs. While expensive observations are needed to detect enough photons to perform a relatively detailed spectral analysis with current instrumentations, \textit{Athena} will be extremely efficient in doing this, thus allowing the gathering of large samples of X-ray detected Hot DOGs with good photon counting statistics. 

\end{enumerate}

\section*{Acknowledgments} \vspace{0.2cm}
We thank the anonymous referee for useful comments and suggestions, which helped in significantly improving this work. FV, WNB, and CTC acknowledge support from the Penn State ACIS Instrument Team Contract SV4-74018 (issued by the Chandra X-ray Center, which is operated by the Smithsonian Astrophysical Observatory for and on behalf of NASA under contract NAS8-03060) and the V.M. Willaman Endowment. RJA was supported by FONDECYT grant number 1151408. PE and DS acknowledge support from Chandra Award Number 17700696 issued by the Chandra X-ray Center. DJW acknowledges support from STFC in the form of an Ernest Rutherford Fellowship. J.W. acknowledges support from MSTC through grant 2016YFA0400702. The Guaranteed Time Observations (GTO) included here were selected by the ACIS Instrument Principal Investigator, Gordon P. Garmire, of the Huntingdon Institute for X-ray Astronomy, LLC, which is under contract to the Smithsonian Astrophysical Observatory; Contract SV2-82024. This publication makes use of data products from the Wide-field Infrared Survey Explorer, which is a joint project of the University of California, Los Angeles, and the Jet Propulsion Laboratory/California Institute of Technology, funded by the National Aeronautics and Space Administration, and was supported by NASA under Proposal No. 13-ADAP13-0092 issued through the Astrophysics Data Analysis Program. Based on observations obtained with XMM-Newton, an ESA science mission with instruments and contributions directly funded by ESA  Member States and NASA.
\\
\bibliography{biblio}

%
%
%
 
%

%

\appendix
\section{W2026+0716}\label{appendix}
W2026+0716 (RA=20:26:15.18, DEC=+07:16:24.0) is a Hot DOG at $z=2.542$ (optically classified as AGN; Eisenhardt et al., in prep.) and was observed for 25.5 ks (PN exposure time, filtered for background flares) as part of the same \xmm\, program targeting W1838+3429 and W2246--0526. However, its bolometric luminosity (\mbox{$L_{\mathrm{bol}}=5\times10^{13}\,L_\odot$}, computed  similarly to the $L_\mathrm{bol}$ values in \citealt{Tsai15}; see Tsai et al., in prep.) is lower than for the other objects we considered in this work ($L_{\mathrm{bol}}\gtrsim10^{14}\,L_\odot$). Therefore, we did not include it in our main analysis, but we considered it when discussing other $L_{\mathrm{bol}}<10^{14}\,L_\odot$ Hot DOGs (e.g., Fig.~\ref{L_NH}--\ref{LX_Lbol}). Its monochromatic luminosity at $6\,\mu m$ is $L_{6\mu m}=1.7\times10^{13}\,L_\odot$ (Tsai et al., in prep.).

W2026+0716 is undetected in both the soft and hard bands, although somewhat enhanced emission can be visually identified in the hard band, suggesting that this source is heavily obscured.
Following the procedure applied for W1838+3429 and W2246--0526 in \S~\ref{results_indiv}, we computed upper limits (at 90\% confidence level) on its photometry, fluxes, and luminosity, assuming the MYTorus model and an obscuring column density of log$N_H\,[\nhunits]=24$. The net-count numbers in the soft and hard energy bands are $<7.7$ and $<22.0$, respectively. Taking into account the proper response matrix and ancillary file, these values correspond to observed fluxes \mbox{$F_{0.5-2\,\mathrm{keV}}<8.69\times10^{-16}\,\funit$} and \mbox{$F_{2-10\,\mathrm{keV}}<1.20\times10^{-14}\,\funit$}, and an intrinsic luminosity \mbox{$L_{2-10\,\mathrm{keV}}<1.62\times10^{45}\,\lunit$}. In particular, the upper limit on luminosity is in agreement with the results we found for our sample of more luminous Hot DOGs.

\end{document}